\documentclass{JHEP3}\newcommand{\format} {\JHEPformat}

\usepackage{amsmath}
\usepackage{epsfig}
\usepackage{latexsym}

\newcommand{\JHEPformat} {
\bibliographystyle{JHEP}
\newcommand{\maketitlepage} {}
\abstract{\theabstract}
\keywords{\thekeywords}
\preprint{\thepreprint}
}

\newcommand{\TITLE}[1] {\newcommand{\thetitle} {#1}\title{#1}}
\newcommand{\ABSTRACT}[1] {\newcommand{\theabstract} {#1}}
\newcommand{\AUTHOR}[1] {\newcommand{\theauthor} {#1}}
\newcommand{\ADDRESS}[1] {\newcommand{\theaddress} {#1}}
\newcommand{\DATE}[1] {\newcommand{\thedate} {#1}\date{#1}}
\newcommand{\KEYWORDS}[1] {\newcommand{\thekeywords} {#1}}
\newcommand{\PREPRINT}[1] {\newcommand{\thepreprint} {#1}}

\newcommand{\half}{{\frac{1}{2}}}

\TITLE{On Penrose limit of elliptic branes}

\AUTHOR{{\bf Stanislav Kuperstein}\footnote{\tt kupers@post.tau.ac.il}}

\ADDRESS{School of Physics and Astronomy\\
  The Raymond and Beverly Sackler Faculty of Exact Sciences\\
  Tel Aviv University, Ramat Aviv, 69978, Israel
}

\author{Stanislav Kuperstein\\
School of Physics and Astronomy\\
The Raymond and Beverly Sackler Faculty of Exact Sciences\\
Tel Aviv University, Ramat Aviv, 69978, Israel\\
E-mail:
\email{kupers@post.tau.ac.il}
}

\ABSTRACT{

We discuss a Penrose limit of an elliptic brane configuration with $N_1$
NS5 and $N_2$ D4 branes. This background is T-dual to $N_1$ D3 branes
at a fixed point of a $\mathbf{C}^3/\mathbf{Z}_{N_2}$ singularity and
the T-duality survives the Penrose limit. The triple scaling limit of
$N_1$ and $N_2$ gives rise to IIA pp-wave solution with a space-like
compact direction. We identify the quiver gauge theory operators and
argue that upon exchange of the momentum along the compact direction 
and the winding number these operators coincide with the operators derived
in the dual type IIB description. We also find a new Penrose limit of the type
IIB background and the corresponding limit in the type IIA picture. In the
coordinate system we use there are two manifest space-like
isometries. The quiver gauge theory operator duals of
the string states are built of three bosonic fields.
}

\DATE{March 2003}
\PREPRINT{TAUP-2723-03\\{\tt hep-th/0303042}}

\format

\begin{document}

\maketitlepage

\section{Introduction}

Recently, new insight into the AdS/CFT correspondence has been achieved by
considering a Penrose limit of the AdS background of type IIB string theory ~\cite{Berenstein:2002jq}. 
The limit for the $AdS \times S^5$ background provides
 a maximally supersymmetric solvable pp-wave background
 ~\cite{Blau:2001ne,Blau:2002dy,Blau:2002mw,Metsaev:2001bj} with
string states related to a special subset of operators in the dual
$\mathcal{N}=4$ gauge theory. These results were extended to other
conformal and non-conformal backgrounds 
~\cite{Itzhaki:2002kh,Gomis:2002km,PandoZayas:2002rx,
Corrado:2002wi,Hubeny:2002vf,Gimon:2002sf,Brecher:2002ar,Oz:2002ku,
Bhattacharya:2002zf,Bhattacharya:2002qx}. 
In particular, the Penrose
limits of $\mathcal{N}=2$ quiver gauge theories were studied
~\cite{Alishahiha:2002ev,
Mukhi:2002ck,Alishahiha:2002jj,Naculich:2002fh,Bertolini:2002nr}. 
These theories correspond to type IIB $N_1$ $D3$-branes at the fixed point of a 
$\mathbf{C_2}/\mathbf{Z_{N_2}}$ singularity ~\cite{Douglas:1996sw}
 (see ~\cite{Bertolini:2002nr} for more complicated
orbifolds leading to $\mathcal{N}=1$ theories).  Under the Penrose limit
(around a geodesic lying away from the fixed point)
the orbifolding of the original background gives rise to a compact
null ~\cite{Alishahiha:2002ev,Mukhi:2002ck} or a space-like direction ~\cite{Bertolini:2002nr} 
and the related momenta has discrete values.
The BMN operators on the gauge theory side are gauge invariant combinations
made out of bi-fundamental and/or adjoint fields. There is a simple moose/quiver
diagram representation of these operators. 
It turns out that this construction reveals an interesting feature ~\cite{Mukhi:2002ck}: 
the discrete momentum along a compact dimension is related to the 
winding number that the ``string'' of the fields winds around the
moose diagram. Conversely, the gauge theory operators that describe
 string states with a non-zero winding number look very much like
momentum states. 
Similarly to the BMN construction higher states of the supergravity
are realized in the gauge theory by various insertion  into
the ``string'' of fields.

The maximally supersymmetric pp-wave solution has two null-like and
28 space-like Killing vectors.
Remarkably, the aforementioned Penrose limits
~\cite{Alishahiha:2002ev,Mukhi:2002ck,Alishahiha:2002jj,
Bertolini:2002nr} of
the $AdS_5 \times S^5/\mathbf{Z_{N_2}}$  background are the same as the usual pp-wave 
background, except that a circle generated by one of the isometries is compactified.
Moreover, the scaling limit of $N_1$ and $N_2$ depends on the isometry,
we are using ~\cite{Bertolini:2002nr}. In order to get a null-like circle  with a finite radius 
like in ~\cite{Alishahiha:2002ev,Mukhi:2002ck,Alishahiha:2002jj} one
needs to scale $N_1$ as $N_2$ 
(the double scaling limit),
while the space-like circle compactified in ~\cite{Bertolini:2002nr} has 
a finite radius if $N_1$ scales as $N_2^3$ (the triple scaling limit).

Performing T-duality along one of the space-like circles ~\cite{Michelson:2002wa} we get
a type IIA pp-wave configuration. 
The string theory on this background was studied in
~\cite{Alishahiha:2002nf}
 and ~\cite{Mizoguchi:2003be}.
In fact, it was shown in ~\cite{Mizoguchi:2003be}, 
that the T-duality is consistent also quantum mechanically
and the two light-cone Hamiltonians have the same spectra.

In this paper we argue that the type IIA pp-wave background of  
~\cite{Michelson:2002wa,Alishahiha:2002nf} arises
in the Penrose limit of the type IIA elliptic brane configuration ~\cite{Fayyazuddin:1999zu}.
This configuration describes $N_1$ $D4$-branes and $N_2$ $NS5$-branes and is T-dual
to the $AdS_5 \times S^5/\mathbf{Z_{N_2}}$  background.
The fact that the T-duality survives the Penrose limit is not surprising.
The commutation of T-duality and Penrose limit was pointed out in ~\cite{Gueven:2000ru}. 

\FIGURE[b]{
 \label{pict}
\centerline{\begin{picture}(0,0)%
\includegraphics{picture.pstex}%
\end{picture}%
\setlength{\unitlength}{4144sp}%
\begingroup\makeatletter\ifx\SetFigFont\undefined%
\gdef\SetFigFont#1#2#3#4#5{%
  \reset@font\fontsize{#1}{#2pt}%
  \fontfamily{#3}\fontseries{#4}\fontshape{#5}%
  \selectfont}%
\fi\endgroup%
\begin{picture}(6322,3127)(2701,-6104)
\put(5716,-4651){\makebox(0,0)[lb]{\smash{{\SetFigFont{12}{14.4}{\rmdefault}{\mddefault}{\updefault}T-duality}}}}
\put(4996,-6046){\makebox(0,0)[lb]{\smash{{\SetFigFont{12}{14.4}{\rmdefault}{\mddefault}{\updefault}The ground state is $\left| p^+; m,k \right>$.}}}}
\put(4996,-5821){\makebox(0,0)[lb]{\smash{{\SetFigFont{12}{14.4}{\rmdefault}{\mddefault}{\updefault}pp-wave background.}}}}
\put(4996,-5641){\makebox(0,0)[lb]{\smash{{\SetFigFont{12}{14.4}{\rmdefault}{\mddefault}{\updefault}The type IIA}}}}
\put(4996,-3751){\makebox(0,0)[lb]{\smash{{\SetFigFont{12}{14.4}{\rmdefault}{\mddefault}{\updefault}$\left |p^+; k,m \right>$.}}}}
\put(4996,-3526){\makebox(0,0)[lb]{\smash{{\SetFigFont{12}{14.4}{\rmdefault}{\mddefault}{\updefault}The ground state is}}}}
\put(2701,-3706){\makebox(0,0)[lb]{\smash{{\SetFigFont{12}{14.4}{\rmdefault}{\mddefault}{\updefault}$AdS_5 \times S_5/\mathbf{Z}_{N_2}$}}}}
\put(7516,-5326){\makebox(0,0)[lb]{\smash{{\SetFigFont{12}{14.4}{\rmdefault}{\mddefault}{\updefault}"momentum" $m$.}}}}
\put(7516,-5101){\makebox(0,0)[lb]{\smash{{\SetFigFont{12}{14.4}{\rmdefault}{\mddefault}{\updefault}number $k$ and}}}}
\put(7516,-4876){\makebox(0,0)[lb]{\smash{{\SetFigFont{12}{14.4}{\rmdefault}{\mddefault}{\updefault}state has winding}}}}
\put(7516,-4651){\makebox(0,0)[lb]{\smash{{\SetFigFont{12}{14.4}{\rmdefault}{\mddefault}{\updefault}dual to the ground}}}}
\put(7516,-4426){\makebox(0,0)[lb]{\smash{{\SetFigFont{12}{14.4}{\rmdefault}{\mddefault}{\updefault}The BMN operator }}}}
\put(7516,-4201){\makebox(0,0)[lb]{\smash{{\SetFigFont{12}{14.4}{\rmdefault}{\mddefault}{\updefault}dual gauge theory.}}}}
\put(7516,-3976){\makebox(0,0)[lb]{\smash{{\SetFigFont{12}{14.4}{\rmdefault}{\mddefault}{\updefault}A subsector of the}}}}
\put(4996,-3301){\makebox(0,0)[lb]{\smash{{\SetFigFont{12}{14.4}{\rmdefault}{\mddefault}{\updefault}pp-wave background.}}}}
\put(2701,-6046){\makebox(0,0)[lb]{\smash{{\SetFigFont{12}{14.4}{\rmdefault}{\mddefault}{\updefault}configuration}}}}
\put(2701,-5596){\makebox(0,0)[lb]{\smash{{\SetFigFont{12}{14.4}{\rmdefault}{\mddefault}{\updefault}The type IIA}}}}
\put(4996,-3121){\makebox(0,0)[lb]{\smash{{\SetFigFont{12}{14.4}{\rmdefault}{\mddefault}{\updefault}The type IIb}}}}
\put(3376,-4651){\makebox(0,0)[lb]{\smash{{\SetFigFont{12}{14.4}{\rmdefault}{\mddefault}{\updefault}T-duality}}}}
\put(4276,-5641){\makebox(0,0)[lb]{\smash{{\SetFigFont{12}{14.4}{\rmdefault}{\mddefault}{\updefault}PL}}}}
\put(4276,-3436){\makebox(0,0)[lb]{\smash{{\SetFigFont{12}{14.4}{\rmdefault}{\mddefault}{\updefault}PL}}}}
\put(2701,-5821){\makebox(0,0)[lb]{\smash{{\SetFigFont{12}{14.4}{\rmdefault}{\mddefault}{\updefault}elliptic brane }}}}
\put(2701,-3481){\makebox(0,0)[lb]{\smash{{\SetFigFont{12}{14.4}{\rmdefault}{\mddefault}{\updefault}SUGRA on}}}}
\put(2701,-3256){\makebox(0,0)[lb]{\smash{{\SetFigFont{12}{14.4}{\rmdefault}{\mddefault}{\updefault}The type IIB}}}}
\end{picture}%
}
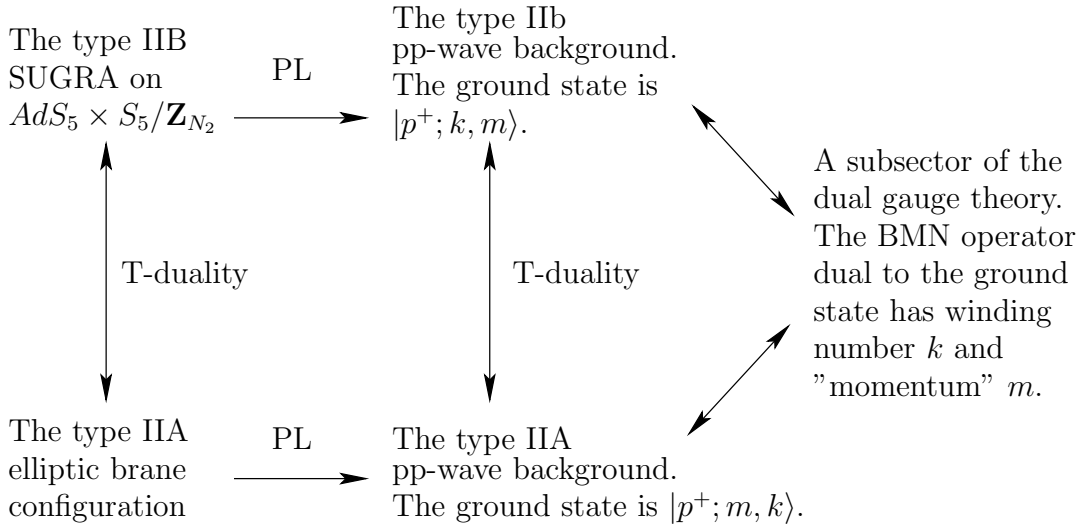
\caption{The T-duality of the type IIB and the type IIA solutions
  commutes with the Penrose limit. In the type IIB pp-wave background string
  states are characterized by the discrete momentum along the compact
  direction $k$ and the winding number $m$ and in the type IIA
  background the roles of $k$ and $m$ are being exchanged.
  In the quiver gauge theory the BMN operator dual to the ground state
  has the same "momentum" and winding number as those of the type IIA
  ground state.}
}

Since the gauge theory dual is the same for these two supergravity solutions, one may expect that
the BMN operators constructed by using the type IIA description of the supergravity are 
``dual'' to the related operators in the type IIB theory. Namely, the 
operator corresponding to a string state with a non-zero discrete momentum 
will look like a momentum operator, while the winding number of the state
will be the winding number that the BMN operator winds around the
moose diagram.
The web of Penrose limits and T-dualities is shown on figure \ref{pict}.

We verify this proposal by taking a Penrose limit of the elliptic brane configuration.
As we already noted the Penrose limit yields the type IIA pp-wave background discussed in
~\cite{Alishahiha:2002nf,Mizoguchi:2003be}.
We calculate the string spectrum including the winding modes around the space-like compact dimension.
 Both the bosonic and the fermionic light-cone 
Hamiltonians appear to be the same as in the T-dual type IIB background ~\cite{Bertolini:2002nr}.
Identifying the $R$-symmetry generators of the gauge theory with the related isometries
of the supergravity solution  
we build the BMN operators related to various string states.
 These operators are essentially the same  
as in ~\cite{Bertolini:2002nr} once we exchange the momentum the the winding number.
One can determine the anomalous dimensions of these operators
using the light-cone Hamiltonian on the supergravity side. The equivalence of the Hamiltonian 
in the type IIA and the type IIB descriptions becomes, therefore, an important consistency check.

We also find a new Penrose limit of the type IIB background and its type IIA counterpart.
The later is distinct from the Penrose limit of
~\cite{Bertolini:2002nr} in that the geodesic does not lie 
along the equator of $S^5$. Due to the shift of the geodesic the pp-wave solution we find
is not the same as in  ~\cite{Bertolini:2002nr}. 
In particular, the radius of the compact dimension is different
and there are two "magnetic" terms in the metric.
Each term gives rise to a massless mode along an appropriate direction
and one of the momenta is quantized.
These magnetic terms also give rise to non-trivial components of the spin connection.
Plugging it together with the RR forms into the Green-Schwarz action
we find a fermionic spectrum with two zero massless modes.
The identification of the BMN operators is also modified.
The operators related to the string ground states are constructed out of three
gauge theory fields with one of them being the adjoint field $\Phi^I$.   
Again, the duality between the string states and the BMN operators found in ~\cite{Mukhi:2002ck} for the
type IIB description does not exist in the T-dual type IIA picture.

In the next section we briefly review the results of
~\cite{Alishahiha:2002ev,Mukhi:2002ck,Alishahiha:2002jj,Bertolini:2002nr}.
Section 3 is devoted to the Penrose limit of the elliptic branes.
The quantum spectrum of the resulting pp-wave background, 
the number of supercharges preserved by it, the compactness of 
various directions and other issues are  discussed.  
In section 4 we construct the field theory operators describing the
string states and compare them to the operators in ~\cite{Bertolini:2002nr}.
In section 5 we introduce a new ``pair'' of Penrose limits and discuss the 
pp-wave/gauge theory duality.

\section{Brief review of the Penrose limits 
 of $AdS_5 \times \mathbf{S}^5/\mathbf{Z}_{N_2}$}

Let us briefly review the results of 
~\cite{Bertolini:2002nr,Mukhi:2002ck,Alishahiha:2002ev,Alishahiha:2002jj}.
In global coordinates the metric of  $AdS_5 \times \mathbf{S}^5/\mathbf{Z}_{N_2}$
can be written as:

\begin{equation}        \label{eq:AdS5S5}
\begin{array}{l}
ds^2 = R^2 \Big[ 
-dt^2 \cosh^2 \rho +  d \rho^2 + \sinh^2 \rho d \Omega_3^2 +  \\
\qquad
d \alpha^2 + \cos^2 \alpha d \beta^2 +      
\sin^2 \alpha \left( d \gamma^2  + \cos^2 \gamma d\chi^2 
                     + \sin^2 \gamma d \eta^2 \right)  \Big],
\end{array}
\end{equation}
where $R = \left( 4 \pi g_s l_s^4 N_1 N_2\right)^{1/4}$ and
the second line is the metric of a $\mathbf{S}^5$ embedded in
a $\mathbf{C}^3$ space orbifolded with a $\mathbf{Z}_{N_2}$ ALE singularity ~\cite{Douglas:1996sw}.
The  RR five form $F^{(5)}$ is the only non-vanishing background field 
in this configuration. The orbifolding of $\mathbf{S^5}$ is achieved 
by requiring a combined periodicity under

\begin{equation}
\chi \to \chi + \frac{2 \pi}{N_2}, \qquad \eta \to \eta - \frac{2 \pi}{N_2}.
\end{equation}
The $\mathbf{C}^3$ coordinates are related to the angles in the following way:

\begin{equation}        \label{eq:ABPhiz1z2z3}
z_1 = R \cos \alpha e^{i \beta}, \quad 
z_2 = R \sin \alpha \cos \gamma e^{i \chi}, \quad
z_3 = R \sin \alpha \sin \gamma e^{i \eta}.
\end{equation}
To obtain this background one places $N_1$ D3-branes at the fixed point of the singular space 
$\mathbf{C}^3/\mathbf{Z}_{N_2}$. 

There are two distinguishable Penrose limits in the problem with a null geodesic 
being based at the fixed point ($\alpha=0$)  
or away from it ($\alpha =\frac{\pi}{2}$).
In the first case one obtains a pp-wave metric with the $\mathbf{Z}_{N_2}$
singularity in its transverse space, while the second case 
leads to the maximally supersymmetric pp-wave solution with eight equal
world-sheet masses and constant RR five form, like in the regular 
$AdS_5 \times S^5$ case with one important difference: the light-like direction
$x^-$ is compact:

\begin{equation}
x^- \to x^- + 2 \pi R_-.
\end{equation}  
The radius $R_- \equiv \frac{R^2}{2 N_2}$ remains finite 
provided the large $R$ limit is a double scaling limit:

\begin{equation}
N_1 \sim \sigma, \quad N_2 \sim \sigma, \quad R^2 \sim \sigma 
\quad \textrm{for} \quad \sigma \to \infty.
\end{equation}
As a consequence there are states with a non-zero winding number $m$
in the string spectrum and the light cone momentum $P^+$ 
is quantized in units of $\frac{1}{2 R_-}$.
On other hand, the light-cone Hamiltonian $H = 2 P^-$ is
not affected by the compactness of $x^-$.
 
The corresponding gauge theory is the $\mathcal{N}=2$ $SU(N_1)^{N_2}$ 
quiver gauge theory (QGT) with
$SU(2)_R \times U(1)_R$ R-symmetry. The $z_1$, $z_2$ and  $z_3$ 
directions of the $\mathbf{C^3}$
are related to the fields $\Phi_I$, $A_I$ and $B_I$ with the former in the adjoint
of $SU(N_1)^{(I)}$ and the other two fields in the bi-fundamental representations 
$(N_1,\bar{N}_1)$ and $(\bar{N}_1,N_1)$.
Identifying the global charges one finds that among these fields only $A^I$ has 
a zero $H=2 P^-$ value, therefore the simplest gauge-invariant QGT operator related  
to the string theory ground state with $m=0$ and one unit of the light-cone momenta
is proportional to:

\begin{equation}
\mathrm{Tr} \left( A^{1}A^{N_2} \ldots A^{N_2} \right) .
\end{equation}
For arbitrary light-cone momentum one plugs $k$  of these string ``bits'' into the trace.
In the moose diagram representation this corresponds to the total string of $A^I$'s  
wrapping the moose exactly $k$ times. Other string modes with $H > 0$ are obtained 
by an appropriate insertion of the various fields along the string of $A^I$'s.

In ~\cite{Mukhi:2002ck,Alishahiha:2002jj} the null geodesic lies at $\gamma =0$, 
the equator of the $\Omega_3$ part in the $\mathbf{S}^5$ metric.
On the contrary, the authors of ~\cite{Bertolini:2002nr} expanded the metric around 
$\gamma = \frac{\pi}{4}$.
Taking $R \to \infty$ together with:

\begin{eqnarray}
t = \mu x^+ + \frac{ x^-}{\mu R^2},     \quad
\chi = \mu x^+ - \frac{x^-}{\mu R^2} + \frac{x_1}{R},   \quad
\eta = \mu x^+ - \frac{x^-}{\mu R^2} - \frac{x_1}{R} ,  \nonumber \\
\gamma = \frac{\pi}{4} - \frac{x_2}{R}, \quad
\alpha = \frac{\pi}{2} - \frac{\tilde{r}}{R}, \quad
\rho = \frac{r}{R}
\end{eqnarray}
they obtained the following pp-wave solution:

\begin{equation} \label{eq:dBmetric}
ds^2 = - 4 dx^+ dx^- -\mu^2 \left(\sum_{I=3}^8 x_I^2 \right) {dx^+}^2 +
 \sum_{i=1}^8 {dx_i}^2 + 4 \mu x_2 dx_1 dx^+ ,
\end{equation}
where $\sum_{j=5}^8 x_j^2 = r^2$ and $x_3^2 + x_4^2 = \tilde{r}^2$ and the RR five-form is:

\begin{equation}
F^{(5)} = 2 \mu \left( dx_1 \wedge dx_2 \wedge dx_3 \wedge dx_4 
           + dx_5 \wedge dx_5 \wedge dx_6 \wedge dx_7  \right) \wedge dx^+.
\end{equation}
The combined periodicity of $\chi$ and $\eta$ implies that:

\begin{equation}
x_1 \equiv x_1 + 2 \pi \frac{R}{N_2}.
\end{equation}
Recalling the definition of $R$ we see that in the triple scaling limit ~\cite{Bertolini:2002nr}

\begin{equation}       \label{eq:tripleSL}
N_1 \sim \sigma, \quad N_2 \sim \sigma^3, \quad R^2 \sim \sigma 
\quad \textrm{for} \quad \sigma \to \infty
\end{equation}
the direction $x_1$ is compact. 
As we already noted, this background is the same as the usual pp-wave
solution,
though one of the directions is compact. 
To make this statement clear it is useful to change the coordinates:

\begin{equation}
Z^+ = x^+, \quad Z^- = x^- +  \mu x_1 x_2, \quad  
Z_I = x_I \quad \textrm{for} \quad I=3, \cdots, 8 \quad
\end{equation}
and

\begin{equation}
\left( \begin{array}{c} Z_1 \\ Z_2  \end{array}  \right) = 
\left( \begin{array}{cc} \cos(\mu x^+)  & \sin(\mu x^+) \\
                         -\sin(\mu x^+) & \cos(\mu x^+) \end{array}  \right)
\left( \begin{array}{c} x_1 \\ x_2  \end{array}  \right).
\end{equation}
The light-cone Hamiltonian, the light-cone momentum 
and the momentum along the compact direction are

\begin{equation}
H = \frac{2}{\mu} P_- = \Delta - 2 J_R, \quad
P^+ = \frac{\Delta + 2 J_R}{2 \mu R^2}, \quad
P_1 = \frac{2 J_L}{R} , 
\end{equation}
where

\begin{equation}
\Delta = i \partial_t, \quad
J_L = -\frac{i}{2} \left( \partial_{\chi} - \partial_{\eta} \right) \quad
\textrm{and} \quad
J_R = -\frac{i}{2} \left( \partial_{\chi} + \partial_{\eta} \right).
\end{equation}
In particular, the Hamiltonian vanishes for the $A$ and the $B$ field, 
therefore the BMN operators related to the $H=0$ ground state are constructed out of both the
$A^I$'s  and the $B^I$'s. The difference between the number of $A$'s
and $B$'s  in string bits is fixed by the quantized momentum along the compact dimension. 
Given that the $A^I$'s transform in the bi-fundamental representations $(N^I, \bar{N}^I)$ and 
the $B^I$'s - in the $(\bar{N}^I, N^I)$, we conclude again that the value of 
the quantized momentum on the supergravity side is related 
to the wrapping of the moose diagram on the dual QGT side. 
We will come back to these operators, while discussing the BMN operators 
in the type IIA description.

Before proceeding it is worth reviewing the string theory quantization 
in the background (\ref{eq:dBmetric}).
In the light-cone gauge $x^+=\tau$ the bosonic part of the light-cone action is:

\begin{equation}
 S_B = \frac{1}{4\pi l_s^2} \int d \tau  \int_{0}^{2 \pi l_s^2 p^+} d \sigma
 \left[ \sum_{i=1}^8 (\partial_\alpha x_i)^2  +
 \mu^2 \sum_{I=3}^8 x_I^2 + 4 \mu x_2 \partial_\sigma x_1   \right]
\end{equation}
The solution of the equations of motion of $x^I$
($I=3 \ldots 8$) 
are the usual expressions one obtain in the case of the maximally supersymmetric pp-wave:

\begin{equation}
\begin{aligned}
x^i(\tau, \sigma)& = \cos\left(\mu \tau \right) \bar{x}^i + 
                \frac{\sin\left(\mu \tau \right)}{\mu p^+}  p_i +   \\
&  \frac{i}{\sqrt{2 p^+}} 
  \sum_{n \neq 0} \frac{1}{\sqrt{\omega_n}} 
  \left(
  a_n^i e^{-i \left( \omega_n \tau + \frac{n}{l_s^2 p^+} \sigma \right)} -
  {a_n^i}^\dagger e^{i \left( \omega_n \tau + \frac{n}{l_s^2 p^+}  \sigma \right)}
  \right) ,
\end{aligned}
\end{equation}
where

\begin{equation}
\omega_n = \sqrt{\mu^2 + \frac{n^2}{l_s^4 (p^+)^2}}.
\end{equation}
The equations of motion of $x_{1,2}(\tau,\sigma)$ read:

\begin{equation}
\partial^\alpha \partial_\alpha x + 2 \mu i \partial_\tau x = 0,
\end{equation}
where $x(\tau,\sigma) \equiv x_1(\tau,\sigma) + i x_2(\tau,\sigma)$. Solving this equation
and allowing winding
modes around the compact direction $x_1$ we get:

\begin{equation}
\begin{aligned}
x(\tau,\sigma) &= \bar{x} + \frac{i}{2\mu p^+} p + \frac{R_1 m}{l_s^2 p^+} \sigma
               + i \frac{1}{(\mu p^+)^{1/2}} e^{2 i \mu \tau} a_0 + \\
 & + \frac{i}{\sqrt{p^+}} e^{i \mu \tau}
  \sum_{n \neq 0} \frac{1}{\sqrt{\omega_n}} 
  \left(
  a_n e^{i \left( \omega_n \tau + \frac{n}{l_s^2 p^+} \sigma \right)} +
  {\tilde{a}_n}^\dagger e^{-i \left( \omega_n \tau + \frac{n}{l_s^2 p^+}  \sigma \right)}
  \right) 
\end{aligned}                      
\end{equation}

The conjugate momenta of $x_i$'s are:

\begin{equation}
\pi^I = \frac{\dot{x}^I}{2 \pi l_s^2}, \quad
\pi^1 = \frac{\dot{x}^1 + 2 \mu x^2}{2 \pi l_s^2}, \quad
\pi^2 = \frac{\dot{x}^2}{2 \pi l_s^2}.
\end{equation}
Imposing the commutation relations between $x^i$'s and $\pi^i$'s we
obtain the following relations for the oscillators:

\begin{equation}  \label{eq:IIBcr}
\left[ x, p \right] = i, \quad
\left[ a_0, a_0^\dagger \right] = 1, \quad
\left[ a^I_n, {a^I_m}^\dagger \right] = \delta_{nm}, \quad
\left[ a_n, {a_m}^\dagger \right] = \delta_{nm}, \quad
\left[ \tilde{a}_n, {\tilde{a}_m}^\dagger \right] = \delta_{nm}
\end{equation}
and the bosonic part of the Hamiltonian reads:

\begin{equation}
H^B = \frac{m^2 R_1^2}{2 l_s^4 p^+} + 2 \mu N_0 + 
\sum_{n \neq 0}  (\omega_n + \mu) N_n  +
\sum_{n \neq 0}  (\omega_n - \mu) \tilde{N}_n +
\sum_{I=3}^8 \sum_{n = -\infty}^{\infty} \omega_n N_n^{I},
\end{equation}
where

\begin{equation}
N^I_n = a_n^I {a_n^I}^\dagger, \quad
N_n = a_n {a_n}^\dagger, \quad
\tilde{N}_n = \tilde{a}_n {\tilde{a}_n}^\dagger, \quad
N_0 = a_0 {a_0}^\dagger.
\end{equation}
The fermionic part of the Hamiltonian is:

\begin{equation}  \label{eq:IIBlcHF}
H^F = \sum_{n=-\infty}^{\infty}
        \left[
           \sum_{a=1}^{4} \left( \omega_n - \frac{\mu}{2} \right) F^{(a)}_n +
           \sum_{a=5}^{8} \left( \omega_n + \frac{\mu}{2} \right) F^{(a)}_n
        \right] , 
\end{equation}
and we refer the reader to ~\cite{Bertolini:2002nr}
for a detailed derivation of the fermionic spectrum. We will perform similar calculations, 
while deriving the fermionic spectrum of the IIA theory in the next section.

An important comment is in order. We see that there is no quantized momenta
in the expression for the compact boson $x_1$. Notice, however, 
that the first commutation relation in (\ref{eq:IIBcr}) and the compactness of $x_1$
impose that $p$ is quantized in units of $\frac{1}{R_1}$. 
The corresponding quantum number $k$ amounts to the momentum quantization condition
along the compact direction:

\begin{equation}
\int d \sigma \pi_1 = \frac{k}{R_1}.
\end{equation}
As expected the Hamiltonian
does not depend on $k$ and there is an infinite degeneracy related to
these modes. 
Finally, the constraint coming from 
the world-sheet energy momentum tensor assumes the form:

\begin{equation}
\sum_{n \neq 0} n \left[ N_n + \tilde{N}_n + 
  \sum_{I=3}^8 N_n^I + \sum_{1}^8 F^{(a)}_n    \right] = k m. 
\end{equation}

\section{The Penrose limit of elliptic branes}

An elliptic branes system consists of $N_2$ NS5 branes periodically arranged
in a compact direction and $N_1$ D4 branes stretched between them.
Its near horizon limit  ~\cite{Fayyazuddin:1999zu} is:

\begin{equation}
\begin{array}{l}
ds^2 = R^2 \Big[ 
-dt^2 \cosh^2 \rho +  d \rho^2 + \sinh^2 \rho d \Omega_3^2 +  \\
\qquad
+ d \alpha^2 + \cos^2 \alpha d \theta^2 + 
\frac{1}{4} \sin^2 \alpha ( d \tilde{\theta}^2 + 
    \cos^2 \tilde{\theta} d \phi^2 ) +
\displaystyle{\frac{dY^2}{\sin^2 \alpha}}  \Big]
\end{array}
\end{equation}
with \footnote{our RR 4-form differs by the factor of 2 from ~\cite{Fayyazuddin:1999zu}}
 
\begin{equation}
\begin{array}{l}
H^{(3)} = \half R^2 \cos \tilde{\theta} dY \wedge 
                                  d \tilde{\theta} \wedge d \phi 
\qquad
e^\Phi = g_s \frac{N_2}{R} \displaystyle{\frac{1}{\sin \alpha}}          \\
F^{(4)} = -4 \pi  N_1  \cos \alpha \sin^3 \alpha \cos \tilde{\theta} 
             d \alpha \wedge d \theta \wedge  d \tilde{\theta} \wedge d \phi,
\end{array}
\end{equation}
where in our notations the period of $Y$ is 
$2 \pi \left( \frac{N_2}{4 \pi g_s N_1} \right)^{1/2}$.
This supergravity solution can be trusted only away from $\alpha=0$, 
where the dilaton is small.

Following  BMN we search for a geodesic at $\rho=0$. Moreover, we restrict 
ourself to a geodesic with $\alpha = \frac{\pi}{2}$ leaving other possibilities 
to the upcoming sections.
Taking the $R \to \infty$ limit we use the following coordinates:

\begin{equation}
\begin{aligned}
 \alpha & = \frac{\pi}{2} - \frac{x}{R},\quad
t = \mu x^+ + \frac{2 x^-}{\mu R^2}, \quad
\rho = \frac{r}{R}, \quad
\tilde{\theta} = \frac{2 x_2}{R}, \quad    \\
 \half \phi & =  \mu x^+ , \quad
\textrm{and} \quad Y = \frac{x_1}{R},
\end{aligned}
\end{equation}
In the PL the metric takes the form:

\begin{equation}        \label{eq:IIAm1}
ds^2  = -4 dx^- dx^+  
- \mu^2 \left( 4x_2^2 + x_3^2 + x_4^2 + \sum_{I=5}^8 x_I^2 \right) d{x^+}^2  
 + dx_1^2 + dx_2^2 + dx_3^2 + dx_4^2 +\sum_{I=5}^8 dx_I^2 
\end{equation}
and the background fields are:

\begin{equation}         \label{eq:IIAm1f}
H^{(3)}_{+12} = 2 \mu, \quad B^{(2)}_{+1} = 2 \mu x_2,
\quad 
F^{(4)}_{+234} = - 4 \pi \frac{N_1}{R^3} \cdot 4 \mu,
\quad
e^{\Phi} = g_s \frac{N_2}{R}.
\end{equation}
This background is exactly the T-dual of the type IIB
solution (\ref{eq:dBmetric}) we discussed in the previous section. 
This is just as well, since 
the elliptic brane configuration at hand and the background considered by ~\cite{Bertolini:2002nr}
are T-dual and according to ~\cite{Gueven:2000ru} the 
Penrose limit is expected to commute with the T-duality. 
Note, however, that this expectation fails to be true once we consider a T-duality along
a null-like circle. In particular, it is not clear to us what is the T-dual counterpart
of the Penrose limit presented in ~\cite{Mukhi:2002ck}.

The following comments are in order:

\begin{itemize}
\item[$\bullet$]
In order to keep both the dilaton and the 4-form finite as $R \to \infty$
we have to impose the same triple scaling limit as in the PL of type IIB background:

\begin{equation}       \label{eq:tsl}  
N_1  \sim \sigma^3, \quad N_2 \sim \sigma , \quad R \sim \sigma \quad
\textrm{for} \quad \sigma \to \infty.
\end{equation}
We will return to this limit in the next section discussing the QGT coupling constant.
\item[$\bullet$]
We easily verify that the only non-trivial equation of motion:

\begin{equation}
R_{++} = \frac{3}{2 \cdot 3!} H_{+ij} H_+^{ij} +
 e^{2 \Phi} \frac{4}{2 \cdot 4!} F_{+ijk} F_+^{ijk}
\end{equation}
is satisfied. Indeed, the 3-form and the 4-form contributions are $2 \mu^2$ and $8 \mu^2$
respectively, therefore 
$R_{++} = ( 1 \cdot 4 + 1 + 1 + 4 ) \mu^2 = 10 \mu^2$ 
matches perfectly with the forms.

\item[$\bullet$]
The only massless space like direction $x_1$ is compact
in the triple scaling limit with the compactification radius

\begin{equation}
R_1 = \frac{N_2}{R}l_s^2
\end{equation}
This is exactly the value predicted by the T-duality transformation.

\item[$\bullet$]
This background preserves 24 supercharges ~\cite{Michelson:2002wa}. 
To see that consider the dilatino variation.
Since the only non-vanishing fields are $H^{(3)}$ and $F^{(4)}$ we
have:

\begin{equation}
\begin{array}{l}
 \delta \lambda = - \frac{1}{2} H \!\!\!\! /  \sigma^3 \epsilon +
   \frac{1}{4} e^\phi F^{(4)} \!\!\!\!\!\!\!\!\! / \,\,\,\,\, \sigma^1 \epsilon =
    \\
\qquad
= \mu \left( -  \Gamma^{+12} \sigma^3 +  \Gamma^{+234} \sigma^1 \right) \epsilon  =
\frac{1}{2} \mu \Gamma_- \Gamma_{12} \left( 1 + \Gamma_{134} (i \sigma^2) \right) \epsilon
\end{array}
\end{equation}
We see that there are two projection
operators acting on $\epsilon$. Therefore, there are $16 + 8 = 24$ preserved supercharges
(see ~\cite{Bena:2002kq} for a related discussion). 
\end{itemize}

\subsection{The string spectrum of the pp-wave background}
\subsubsection{Bosonic sector}

Now we will work out the spectrum of strings in the background (\ref{eq:IIAm1},\ref{eq:IIAm1f}).
In the light cone gauge $x^+ = \tau$ the bosonic part of the Lagrangian is:

\begin{equation}
- 2 \pi l_s^2 \mathcal{L}^B = \half \left[ \sum_{i=1}^8 (\partial_\alpha x_i)^2 +
 \mu^2 ( \sum_{I=3}^8 x_I^2 + 4 x_2^2 ) \right] + 2 \mu x_2 \partial_\sigma x_1,
\end{equation}
where the last term is the contribution of the constant magnetic field. 

The solution of the equations of motion of $x^I$ for  $I = 3, \ldots, 8$ 
and the related quantization is the same as in the type IIB theory
and we will skip it here. 
The equations of motion of $x_{1,2}$ read:

\begin{equation}
\left \{
\begin{array}{l}
\partial^2_\alpha  x_1 - 2 f  \partial_\sigma x_2 = 0 \\
\partial^2_\alpha  x_2 - 4 f^2 x_2 + 2 f  \partial_\sigma x_1 = 0. 
\end{array}
\right.
\end{equation}
To solve these equations we use the following expansions:

\begin{equation}
\left \{
\begin{array}{rcl}
x_1 &=& \displaystyle{
    \frac{m R_1}{l_s^2 p^+} \sigma +  \frac{i}{\sqrt{2 p^+}} \sum_{n = -\infty}^{\infty}
                       \left( \beta_n(\tau) e^{i n \sigma} -
                           {\beta_n(\tau)}^{\dagger} e^{-i n \sigma} \right) }\\
x_2 &=& \displaystyle{
    \frac{i}{\sqrt{2 p^+}} \sum_{n = -\infty}^{\infty}
                       \left( \gamma_n(\tau) e^{ i n \sigma} -
                           {\gamma_n(\tau)}^{\dagger} e^{- i n \sigma} \right). }\\
\end{array}
\right.
\end{equation}
Thus the equations for $\beta_{n}(\tau)$ and $\gamma_n(\tau)$ assume the form:

\begin{equation}
\begin{aligned}
(& \beta_n^{''}(\tau) - {\beta_{-n}^\dagger}^{''}(\tau)) 
   + n^2 (\beta_n(\tau) - {\beta_{-n}^\dagger}(\tau))
   + \frac{2 i \mu n}{l_s^2 p^+}  (\gamma_n(\tau) - \gamma_{-n}(\tau)) = 0 \\
(& \gamma_n^{''}(\tau) - {\gamma_{-n}^\dagger}^{''}(\tau)) 
   + (n^2+4\mu^2) (\gamma_n(\tau) - {\gamma_{-n}^\dagger}(\tau))
   - \frac{2 i \mu n}{l_s^2 p^+}  (\beta_n(\tau) - \beta_{-n}(\tau)) = 0
\end{aligned}
\end{equation}
The last identity is valid only for $n \neq 0$ and the correct expression
 for $n = 0$ is:

\begin{equation}    \label{eq:gamma0}
\frac{i}{\sqrt{2 p^+}} \left[
(\gamma_0^{''}(\tau) - {\gamma_{0}^\dagger}^{''}(\tau)) 
   + 4 \mu^2 (\gamma_0(\tau) - {\gamma_{0}^\dagger}(\tau)) \right]
   + 2 \mu \frac{m R_1}{l_s^2 p^+}  = 0. 
\end{equation}
We look for a solution in a form:

\begin{equation}
\beta_{n}(\tau) = b_n e^{-i \omega_n \tau}.
\end{equation}
From the equation for $x_1(\tau, \sigma)$ one obtains the expression for $\gamma_n(\tau)$

\begin{equation}
\gamma_{n}(\tau) = - i \frac{l_s^2 p^+}{2 \mu n} 
          \left( \omega_n^2 - \left( \frac{n}{l_s^2 p^+} \right)^2 \right)
          b_n e^{-i \nu_n \tau} \quad \textrm{for} \quad n \neq 0
\end{equation}
while the equation for $x_2(\tau, \sigma)$ leads to a quadratic equation for $\omega_n$
with roots given by:

\begin{equation}
\omega^1_n =  \mu + \sqrt{\mu^2 + \left( \frac{n}{l_s^2 p^+} \right)^2 }  \qquad
\omega^2_n = -\mu + \sqrt{\mu^2 + \left( \frac{n}{l_s^2 p^+} \right)^2 }  
\end{equation}
Notice, that $\omega_{n}^2 $ vanishes for $n=0$. This is a remnant of the translational invariance
along $x_1$. 
 The expression for 
$x_1(\tau,\sigma)$ takes the following form:

\begin{equation}
\begin{aligned}
x_1(\tau,\sigma) =
    \frac{m R_1}{l_s^2 p^+} \sigma + 
     \bar{x}_1 + \frac{p_1}{p^+} \tau + 
    \frac{i}{2 \sqrt{p^+}} \sum_{n \neq 0}  
        \frac{1}{\omega_n^{1/2}}  
                            \Bigg  [
              \left( \frac{\omega^2_n}{\omega^1_n}\right)^{\half} 
                a^1_n e^{-i \left(\omega^1_n \tau - \frac{n}{l_s^2 p^+} \sigma \right)} + \\
              +\left( \frac{\omega^1_n}{\omega^2_n}\right)^{\half}
                {a^2_n}^\dagger e^{-i \left(\omega^2_n \tau - \frac{n}{l_s^2 p^+} \sigma \right)} 
                      - \textrm{c.c.} 
                             \Bigg ]
\end{aligned}
\end{equation}
In order to write the result for $x_2(\tau,\sigma)$ we have to solve (\ref{eq:gamma0}):

\begin{equation}
\frac{i}{\sqrt{2 p^+}} \left[ \gamma_0 (\tau) - {\gamma_{0}^\dagger}(\tau) \right]
 = \frac{i}{2 \sqrt{\mu p^+}}  \left( e^{-2 i \mu \tau} a_0 +  e^{2 i \mu \tau} \bar{a}_0  \right)
 - \frac{m R_1}{2 \mu l_s^2 p^+}.
\end{equation}
Therefore:

\begin{equation}
\begin{array}{l}
x_2(\tau,\sigma) = \displaystyle{
     - \frac{m R_1}{2 \mu l_s^2 p^+}  + 
   \frac{i}{2 \sqrt{\mu p^+}}  
              \left( e^{-2 i \mu \tau} a_0 +  e^{2 i \mu \tau} \bar{a}_0  \right)}+ \\
\qquad  \displaystyle{
       +  \frac{1}{2 \sqrt{p^+}} \sum_{n \neq 0}
      \frac{1}{\omega_n^{1/2}} \left( 
               a^1_n e^{-i \left(\omega^1_n \tau - \frac{n}{l_s^2 p^+} \sigma \right)} + 
               a^2_n e^{-i \left(\omega^2_n \tau - \frac{n}{l_s^2 p^+} \sigma \right)} 
                      + \textrm{c.c.} \right) }
\end{array}
\end{equation}
The conjugate momenta of $x_1$ and $x_2$ are:

\begin{equation}
\Pi^1 = \frac{\dot{x}_1}{2 \pi l_s^2} \qquad \Pi^2 = \frac{\dot{x}_2}{2 \pi l_s^2}  
\end{equation}
and the quantum commutators are

\begin{equation}
\left[ x_1(\tau,\sigma), \Pi^1 (\tau,\sigma^{'}) \right] = i \delta(\sigma -\sigma^{'})
\qquad \left[ x_2(\tau,\sigma), \Pi^2 (\tau,\sigma^{'}) \right] = i \delta(\sigma -\sigma^{'}).
\end{equation}
Plugging the results for $x_1$ and $x_2$ one obtains the following 
relations for the oscillators:

\begin{equation}
\left[ \bar{x}_1, p_1 \right] = i, 
\qquad 
\left[ a^1_n, {a^1_m}^\dagger \right] =  \delta_{nm}, 
\qquad  \left[ a^2_n, {a^2_m}^\dagger \right] = \delta_{nm} 
\qquad \textrm{and} \quad \left[ a_0, {\bar{a}_0} \right] = 1.
\end{equation}
The bosonic part of the Hamiltonian is:

\begin{equation}
\begin{array}{l}
\displaystyle{
H^B = \int d \sigma \left( \Pi_i \dot{x}_i - \mathcal{L}_B \right) =  }\\
\displaystyle{
\qquad \frac{1}{4 \pi l_s^2} \int_{0}^{2 \pi l_s^2 p^+} d \sigma 
   \left[ \sum_{i=1}^8 ({\dot{x}_i}^2 + {{x}_i^{\prime}}^2 ) + 
       \mu^2 \left( \sum_{I=3}^8 x_I^2 + 4 x_2^2 \right) + 4 \mu x_2 \partial_\sigma x_1  \right]}
\end{array}
\end{equation}
or using the mode expansions and the commutators:

\begin{equation}
H^B = \frac{p_1^2}{2 p^+} + 2 \mu N_0 + 
\sum_{n \neq 0}  \omega^1_n N^1_n  +
\sum_{n \neq 0}  \omega^2_n N^2_n +
\sum_{I=3}^8 \sum_{n = -\infty}^{\infty} \omega_n N_n^{I},
\end{equation}
where

\begin{equation}
N^I_n = a_n^I {a_n^I}^\dagger, \quad
N^1_n = a^1_n {a^1_n}^\dagger, \quad
N^2_n = a^2_n {a^2_n}^\dagger, \quad
N_0 = a_0 {\bar{a}_0}.
\end{equation}
Remarkably, this Hamiltonian is exactly the same as the light-cone Hamiltonian 
of the type IIB background. Indeed, the commutation relation between $\bar{x}_1$ and $p_1$
and the compactness of the direction $x_1$ imply the following quantum values of $p_1$:

\begin{equation}
p_1 = \frac{k}{R_1} \quad \textrm{for} \quad k \in \mathbf{Z}.
\end{equation} 
Identifying $k$ with the winding number $m$ arising in the type IIB pp-wave solution and 
recalling the relation between the radii of the compact direction 
in the type IIA and the type IIB description we obtain that:

\begin{equation}
H^B_{IIA} = H^B_{IIB}.
\end{equation}

\subsubsection{Fermionic sector}
In the light cone gauge 

\begin{equation}
x^+ = \tau, \quad \Gamma^+ \theta^A =0
\end{equation}
the Green-Schwarz fermionic action is given by:

\begin{equation}   
S^F_{lc} = \frac{i}{4 \pi l_s^2} \int d \tau \int_{0}^{2 \pi l_s^2 p^+} d \sigma
\left[ 
   \left( \eta^{\alpha \beta} \delta_{AB} - \epsilon^{\alpha \beta} \sigma^3_{AB} \right)
   \partial_\alpha x^+ \bar{\theta}^A \Gamma_+ \left( \mathcal{D}_\beta \theta \right)^B 
\right],
\end{equation}
where $\theta^A$, $A=1,2$ are two 10 dimensional spinors with different chiralities and 
$\sigma^i$ are the Pauli's matrices. The generalized covariant derivative, $\mathcal{D}$, 
in the type IIA supergravity reads:

\begin{equation}
\mathcal{D}_\beta = \partial_\beta + \frac{1}{4} \partial_{\beta} x^+
 \left[  \left(  \omega_{\rho \sigma +} - \frac{1}{2} H_{\rho \sigma +} \sigma^3 \right) 
       \Gamma^{\rho \sigma} + \frac{1}{2 \cdot 4!} F_{\lambda \nu \rho \sigma}
          \Gamma^{\lambda \nu \rho \sigma} \sigma^1 \Gamma_+   \right].
\end{equation}
The four and the three-form have the components $F_{+234}$ and $H_{12+}$ turned on and all the 
components of the spin connection $\omega$ vanish for the metric (\ref{eq:IIAm1}).
This reduces the light cone action to:

\begin{equation}
\begin{aligned}
\frac{i}{4 \pi l_s^2} \int d \tau \int_{0}^{2 \pi l_s^2 p^+} d \sigma
\Bigg[
 \bar{\theta}^1 \Gamma^- \partial_+ \theta^1 +
 \bar{\theta}^2 \Gamma^- \partial_+ \theta^2     
 + \frac{\mu}{2} \left(  \bar{\theta}^1 \Gamma^- \Gamma^{12}  \theta^1 -
                         \bar{\theta}^2 \Gamma^- \Gamma^{12}  \theta^2 \right)  \\
 + \mu \left(  \bar{\theta}^1 \Gamma^- \Gamma^{1234}  \theta^2 -
                         \bar{\theta}^2 \Gamma^- \Gamma^{1234} \theta^1 \right)
\Bigg].
\end{aligned}
\end{equation}

Following ~\cite{Bertolini:2002nr} we use the light-cone condition $\Gamma^+ \theta^A = 0$ 
to rewrite the action in terms of the canonical eight-component spinors
$S^{Ab}$, $b=1, \ldots 8$, which are defined by the relation 
$\Gamma^{+-} \theta^{Aa} = 2^{1/4} {p^+}^{-1/2} S^{Aa}$. Now the action reads:

\begin{equation}
S^F_{lc} = \frac{i}{4 \pi l_s^2 p^+} \int d \tau \int_{0}^{2 \pi l_s^2 p^+} d \sigma
\left[ 
   S^1 \left( \partial_+ - \frac{\mu}{2} \gamma^{12} \right) S^1 +
   S^2 \left( \partial_- - \frac{\mu}{2} \gamma^{12} \right) S^2 +
   2 \mu  S^1 \gamma^{234} S^2 
\right],
\end{equation}   
where $\gamma^i$'s are $8 \times 8$ gamma-matrices. The equations of motion are:

\begin{equation}
\left\{
\begin{array}{rcl}
\left( \partial_+ - \frac{\mu}{2} \gamma^{12} \right) S^1 +  \mu \gamma^{234} S^2 &=& 0 \\
\left( \partial_+ - \frac{\mu}{2} \gamma^{12} \right) S^2 -  \mu \gamma^{234} S^1 &=& 0
\end{array}
\right.
\end{equation} 
To solve these equations we define $S^A = e^{\frac{\mu}{2}} \gamma^{12} \Sigma^A$.
The fields $\Sigma^{1,2}$ obey the equations of motion in the usual pp-wave background
 ~\cite{Metsaev:2002re}:

\begin{equation}
\left\{
\begin{array}{rcl}
 \partial_+ \Sigma^1 -  \mu \gamma^{234} \Sigma^2 &=& 0 \\
 \partial_+ \Sigma^2 +  \mu \gamma^{234} \Sigma^1 &=& 0 
\end{array}
\right.
\end{equation} 
Thus the mode expansion of $S^A$'s is:

\begin{equation}
\begin{aligned}
S^1 &= e^{\frac{\mu}{2} \gamma^{12} \tau} 
\left[ \left(
\frac{1}{\sqrt{2}} e^{- i \mu \tau} S_0 -
 \sum_{n \neq 0} c_n e^{-i \omega_n \tau} \left( S_n e^{i \frac{n}{l_s^2 p^+} \sigma}
                  + \frac{\omega_n - n}{\mu} S_{-n} e^{-i \frac{n}{l_s^2 p^+} \sigma} \right)
\right) + \textrm{h.c.} \right]   \\
S^2 &= e^{\frac{\mu}{2} \gamma^{12} \tau} 
\left[ \left(
-\frac{1}{\sqrt{2}} e^{- i \mu \tau} i \gamma^{234}S_0 -  \right. \right. \\
      &  \left. \left.   - i \gamma^{234}
 \sum_{n \neq 0} c_n e^{-i \omega_n \tau} \left( S_n e^{i \frac{n}{l_s^2 p^+} \sigma}
                  - \frac{\omega_n - n}{\mu} S_{-n} e^{-i \frac{n}{l_s^2 p^+} \sigma} \right)
\right) + \textrm{h.c.} \right],
\end{aligned}
\end{equation}
where $c_n = (1 + \frac{\omega_n -n}{\mu})^{-1/2}$.
In terms of $S^1$ and $S^2$ the fermionic part of the light-cone Hamiltonian is:

\begin{equation}
H^F = \frac{i}{4 \pi l_s^2 p^+} \int d \tau \int_{0}^{2 \pi l_s^2 p^+} d \sigma
\left( S^1 \dot{S}^1 +  S^2 \dot{S}^2 \right)
\end{equation}
and under the quantization we end up with

\begin{equation}
H^F = \sum_{n=-\infty}^{\infty} S^\dagger_n (\omega_n + i \frac{\mu}{2} \gamma^{12}) S_n.
\end{equation}
This Hamiltonian is exactly the same as the fermionic Hamiltonian in the type IIB background 
~\cite{Bertolini:2002nr}. In particular, using the fact that the eigenvalues of $i \gamma^{12}$
are $\pm 1$, each with multiplicity four, one can re-write the Hamiltonian  
in the form (\ref{eq:IIBlcHF}).

\section{The BMN operators in the dual QGT}

We now want to relate the string spectrum to states in the dual gauge theory.
At energies below $1/R_1$, where $R_1$ is the radius of the compact dimension,
the gauge theory on the stack of $N_1$ of D4-branes 
is a $3+1$ dimensional $\mathcal{N}=2$$SU(N_1)^{N_2}$ quiver  gauge theory (QGT) with a
$SU(2)_R \times U(1)_R$ $R$-symmetry group. Notice that the radius $R_1$ is kept fixed 
in the triple scaling limit and we may use the four dimensional description.
In 11 dimensions the elliptic branes system we discuss is realized as a M5-M5' branes configuration.
One has $N_1$ M5-branes with world-volume coordinates $(0,1,2,3,4,5)$ and 
$N_2$ M5-branes with world-volume coordinates $(0,1,2,3,6,7)$. 
In this background the $R$-symmetry corresponds to the $SU(2)_{8910} \times U(1)_{45}$ rotational
invariance.
In ten dimensions the $SU(2)$ part of the $R$-symmetry is realized by the
$d \tilde{\theta}^2 + \cos \tilde{\theta}^2 \phi^2$ term in the metric and the $U(1)$
part is related to the rotational invariance along $\theta$ direction. Therefore,
we identify the $SU(2)_R$ Cartan current with $J_\phi=- i \partial_\phi$ and the generator of 
$U(1)_R$ with $J_\theta=- i \partial_\theta$.
The bosonic part of the gauge theory hyper-multiplets consists of two bi-fundamentals 
$A_I$ in the $(N_1,\bar{N}_1)$ of the $SU(N_1)^{(I)} \times SU(N_1)^{(I+1)}$ and
$B_I$ in the $(\bar{N}_1,N_1)$. Moreover, $(A,\bar{B})$ as well as $(\bar{A},B)$
are doublets of $SU(2)_R$ and have zero charges under $U(1)_R$. The third bosonic field 
$\Phi_I$ comes from the vector multiplet  and transforms 
in the adjoint of $SU(N_1)^{{I}}$. This field is a singlet of $SU(2)_{R}$ and 
the $U(1)_R$ factor acts on it as phase rotations.
The adjoint fields $\Phi_I$ can be thought of as describing fluctuations of the D4 branes along 
the NS5 branes. The bi-fundamental fields are realized in the brane description as open strings
stretched along the compact dimension between two adjacent NS5 branes.
Thus, a shift by one period along the comact direction acts on the fields as:

\begin{equation}  \label{eq:Shift}
\Phi^I \to \Phi^I, \qquad A^I \to A^{I+1}, \qquad B^I \to B^{I-1}. 
\end{equation}
In terms of the currents the world-sheet Hamiltonian and the momentum are:

\begin{equation}    \label{eq:dict1}
\begin{array}{l}
\displaystyle{
p^+ = - \half p_- = \half i \partial_{x^-} =  i\frac{\partial_{t}}{\mu R^2}
   =  \frac{\Delta}{\mu R^2}
}      \\
\displaystyle{
H = - p_+ = i \partial_{x^+} =  \mu \left( \Delta - 2 J_\phi \right) + O(R^{-1})
}
\end{array}.
\end{equation}
In particular, the triple scaling limit translates into:

\begin{equation}
\Delta \sim  2 J_\phi \sim R^2 \sim \sigma^2 
   \quad \textrm{for} \quad \sigma \to \infty.
\end{equation}
Following these observations we can summarize the quantum numbers of various bosonic
fields:

\begin{center}
\begin{tabular}{c|c|c|c|c|c|}
& $\Delta$  & $J_\theta$ &  $J_\phi$ & $H$ \\
 \hline
         $A_I$  &  1  &  0  &  $\half$   &  0   \\  \hline  
         $B_I$  &  1  &  0  &  $\half$   &  0   \\  \hline  
   $\bar{A}_I$  &  1  &  0  &  $-\half$  &  2   \\  \hline 
   $\bar{B}_I$  &  1  &  0  &  $-\half$  &  2   \\  \hline  
      $\Phi_I$  &  1  &  1  &     0      &  1   \\  \hline
$\bar{\Phi}_I$  &  1  &  -1 &     0      &  1   \\  \hline 
\end{tabular}
\end{center}
The dictionary between the gauge theory and the string theory
is much similar to the type IIB description ~\cite{Bertolini:2002nr}. 
The only difference is the identification of
the momentum along the compact direction. In the type IIB picture $p_1$ is related
to a generator of a $U(1)$ symmetry, which rotates $A$ and $B$ in opposite directions.
In the type IIA picture the discrete values of $p_1$ are realized in the gauge theory 
as eigenvalues of the shift operator (\ref{eq:Shift}).

We can write the  prediction of the energy eigenvalues from free string
theory using that:

\begin{equation}
\mu p^+ = \frac{\Delta}{R^2} \sim \frac{2 J_\phi}{R^2} = \frac{2 J_\phi}{\sqrt{g^2_{QGT} N_1}},
\end{equation} 
where $g^2_{QGT} = 4 \pi g_s N_2$ is the QGT coupling constant in each gauge group factor.
The expression for the frequencies $\omega_n$'s is:

\begin{equation}
\omega_n = \mu \sqrt{1 + n^2 \frac{g^2_{QGT} N_1}{J^2}}, 
\end{equation}
where $J \equiv 2 J_\phi$. Plugging this result into the light-cone Hamiltonian
 one obtains $H^B$ and $H^F$ completely in terms
of the gauge theory parameters ~\cite{Bertolini:2002nr}.

Notice, that the coupling constant $g_{QGT}$ diverges in the triple scaling limit.
As explained in ~\cite{Berenstein:2002jq} in a slightly different context, due 
to various supersymmetry cancellations the relevant quantity that should be 
kept fixed is precisely $\frac{g^2_{QGT} N_1}{J}$. It is straightforward to check
that this value is indeed finite under our scaling and the prediction for the energy
eigenvalues does make sense.

The ground state in the gauge theory is parameterized by the winding number
$m$ and the momentum $k$ along the compact direction. We will denote it
by $\left|k,m; p^+ \right>$. According to the level matching condition there is no ground state 
with $km \neq 0$. For the later convenience we adopt the following notations:

\begin{equation}
\mathbf{A} = 
\left( 
\begin{array}{ccccc}  
0 & A_1 & 0 & \cdots & 0  \\
0 & 0 & A_2 & \cdots & 0  \\ 
\vdots & & & \ddots  & \vdots \\
0 & 0 & 0 & \cdots   & A_{N_2-1}  \\
A_{N_2} & 0 & 0 & \cdots   & 0
\end{array}
\right)
\qquad
\mathbf{B} = 
\left( 
\begin{array}{ccccc}  
0      & 0 & \cdots &   0        & B_{N_2}  \\
B_1    & 0 & \cdots &   0        & 0 \\ 
0      & 0 & B_2    & \cdots     & 0  \\ 
\vdots &   & \ddots & \vdots     & \vdots \\
0      & 0 & \cdots & B_{N_2-1}  & 0
\end{array}
\right) 
\end{equation}
and

\begin{equation}
\mathbf{\Phi} = 
\left( 
\begin{array}{cccc}  
\Phi_1 & 0      &  \cdots & 0  \\
0      & \Phi_2 &  \cdots & 0  \\ 
\vdots &        & \ddots  & \vdots   \\
0      & 0      & \cdots  & \Phi_{N_2}  
\end{array}
\right),
\end{equation}
where the blocks $A_I,B_I, \Phi_I$ are $N_1 \times N_1$ matrices. We also need 
the shift and the clock matrices:

\begin{equation}
\mathbf{U} = 
\left( 
\begin{array}{ccccc}  
0 & 1 & 0 & \cdots & 0  \\
0 & 0 & 1 & \cdots & 0  \\ 
\vdots & & & \ddots  & \vdots \\
0 & 0 & 0 & \cdots   & 1  \\
1 & 0 & 0 & \cdots   & 0
\end{array}
\right)
\quad \textrm{and} \quad 
\mathbf{V} = 
\left( 
\begin{array}{cccc}  
\theta & 0      &  \cdots & 0  \\
0      & \theta^2 &  \cdots & 0  \\ 
\vdots &        & \ddots  & \vdots   \\
0      & 0      & \cdots  & \theta^{N_2}  
\end{array}
\right),
\quad \theta = e^{i\frac{2 \pi }{N_2}}.
\end{equation}
In this notation the shift operator (\ref{eq:Shift}) acts as:

\begin{equation}
\mathbf{A} \to \mathbf{U}^t \mathbf{A} \mathbf{U}, \quad
\mathbf{B} \to \mathbf{U}^t \mathbf{B} \mathbf{U}, \quad
\textrm{and} \quad
\mathbf{\Phi} \to \mathbf{\Phi}.
\end{equation}
We propose that the BMN operator describing a momentum state of the type IIA 
supergravity is exactly the operator which is related to the winding string state
$\left| k=0, m \right>_{\mathrm{IIB}}$ in the type IIB description. Namely:

\begin{equation}
\left| k, m=0 \right>_{\mathrm{IIA}} 
 \simeq \mathrm{Tr} \left[ 
            \mathbf{V}^k \mathcal{G}_{J,J} \left( \mathbf{A}, \mathbf{B}; \omega^k \right) 
                    \right],
\end{equation}
where $\omega = \theta^{1/J}$ and

\begin{equation}               \label{eq:G}
\displaystyle{
\mathcal{G}_{K,L} \left( A, B; \omega^m \right) \equiv
\frac{1}{\sqrt{(K+L)!K!L!}} \partial_x^K  \partial_y^L
  \prod_{s=0}^{K+L-1} \left( \omega^{-s/2} x \mathbf{A} + 
        \omega^{s/2} y \mathbf{B} \right) \vert_{x=y=0}.  
}
\end{equation}
Indeed, under (\ref{eq:Shift}) the operator transforms as:

\begin{eqnarray}
\mathrm{Tr} 
 \left[ \mathbf{V}^k \mathcal{G}_{J,J}
            \left( \mathbf{U^t A U}, \mathbf{U^t B U}; \omega^k \right) \right]
= \mathrm{Tr} 
 \left[\mathbf{U} \mathbf{V}^k \mathbf{U^t} \mathcal{G}_{J,J} 
                                     \left( \mathbf{A}, \mathbf{B};  \omega^k \right) \right] = \\
= e^{2 \pi i\frac{k}{N_2}} \mathrm{Tr} 
 \left[ \mathbf{V}^k \mathcal{G}_{J,J} \left( \mathbf{A}, \mathbf{B}; \omega^k \right) \right],
\end{eqnarray}
where we used cyclicity invariance of the trace, 
the basic properties of the clock and the shift matrices and the definition of $\theta$.
We further write the ground state with a non-zero winding number:

\begin{equation}
\left| k=0, m \right>_{\mathrm{IIA}}
 \simeq \mathrm{Tr} \left[ 
             \mathcal{G}_{J+\frac{m N_2}{2},J-\frac{m N_2}{2}} 
                \left( \mathbf{A}, \mathbf{B}; 1 \right) 
                    \right].
\end{equation}
In the type IIB description this operator describes the momentum state 
$\left| k, m=0 \right>_{\mathrm{IIB}}$.

The reader can find in ~\cite{Bertolini:2002nr} the full set of the 
BMN operators reproducing the string spectrum as well as the proof of the level matching,
the normalization and the anomalous dimension calculation. Replacing the momentum number $k$
and the winding number $m$ one gets all of the BMN operators corresponding to the string states
in the type IIA background.

\section{A new pair of Penrose limits}
\subsection{IIB}

We start the background (\ref{eq:AdS5S5}) and take the Penrose limit around the geodesic
at $\alpha =\frac{\pi}{4}$. We use the following coordinate transformation

\begin{eqnarray}     \label{eq:CT2IIB}
t = \mu x^+ + \frac{2 x^-}{\mu R^2},     \quad
\chi = \mu x^+  - \frac{x_3}{R} + \sqrt{2}\frac{x_1}{R},   \quad
\eta = \mu x^+  - \frac{x_3}{R} - \sqrt{2}\frac{x_1}{R},   \quad   \nonumber     \\
\beta = \mu x^+ + \frac{x_3}{R}, \quad                 
\gamma = \frac{\pi}{4} - \sqrt{2}\frac{x_2}{R}, \quad
\alpha = \frac{\pi}{4} - \frac{x_4}{R}, \quad
\rho = \frac{r}{R} \qquad.
\end{eqnarray}
In the $R \to \infty$ limit we end up with a pp-wave metric which has two
"magnetic" terms:

\begin{equation}    \label{eq:IIBm2}
ds^2 = - 4 dx^+ dx^- -\mu^2 \left(\sum_{I=5}^8 x_I^2 \right) {dx^+}^2 +
 \sum_{i=1}^8 {dx_i}^2 + 4 \mu x_2 dx_1 dx^+ + 4 \mu x_4 dx_3 dx^+,
\end{equation}
where $r^2=\sum_{I=5}^8 x_I^2$. Similarly to the $\alpha = \frac{\pi}{2}$ case
the combined periodicity of $\chi$ and $\eta$ leads to:

\begin{equation}
x_1 \equiv x_1 + 2 \pi \frac{R}{\sqrt{2} N_2}
\end{equation}
and the compactification radius remains finite 
in the triple scaling limit (\ref{eq:tripleSL}).
One might use the coordinate transformation discussed in section 2 to re-write the metric in 
the conventional pp-wave form. 
One obtains this way eight world-sheet bosonic masses equal to $\mu^2$.
We will not apply this transformation since in new coordinates the isometry along the compact
direction $x_1$ is not manifest. Moreover, we want to keep the "magnetic" term 
$ 4 \mu x_4 dx_3 dx^+$, because the identification of BMN operators 
simplifies in these coordinates as we will see in the next section.

The calculation of the bosonic spectrum follows the steps of section 2.
The equations of motion of $x^{1,2,3,4}(\tau,\sigma)$ are: 

\begin{equation}
\partial^\alpha \partial_\alpha x_{(12)} +  \mu i \partial_\tau x_{(12)} = 0, \qquad
\partial^\alpha \partial_\alpha x_{(34)} +  \mu i \partial_\tau x_{(34)} = 0,
\end{equation}
where  $x_{(12)}(\tau,\sigma) \equiv x_1(\tau,\sigma) + i x_2(\tau,\sigma)$ 
and $x_{(34)}(\tau,\sigma) \equiv x_3(\tau,\sigma) + i x_4(\tau,\sigma)$.
The solutions are:

\begin{eqnarray}
x_{(12)}(\tau,\sigma) &=& \bar{x}_1 + \frac{i}{2\mu p^+} p_2 + 
                    \frac{R_1^{\mathbf{IIB}} m}{l_s^2 p^+} \sigma
               + i \frac{1}{(\mu p^+)^{1/2}} e^{2 i \mu \tau} a_0 +   \nonumber\\
 && + \frac{i}{\sqrt{p^+}} e^{i \mu \tau}
  \sum_{n \neq 0} \frac{1}{\sqrt{\omega_n}} 
  \left(
  a_n e^{i \left( \omega_n \tau + \frac{n}{l_s^2 p^+} \sigma \right)} +
  {\tilde{a}_n}^\dagger e^{-i \left( \omega_n \tau + \frac{n}{l_s^2 p^+}  \sigma \right)}
  \right)     
\end{eqnarray}
and 
\begin{eqnarray}
x_{(34)}(\tau,\sigma) &=& \bar{x}_3 + \frac{i}{2\mu p^+} p_4 
               + i \frac{1}{(\mu p^+)^{1/2}} e^{2 i \mu \tau} a_0 +  \nonumber\\
 && + \frac{i}{\sqrt{p^+}} e^{i \mu \tau}
  \sum_{n \neq 0} \frac{1}{\sqrt{\omega_n}} 
  \left(
  b_n e^{i \left( \omega_n \tau + \frac{n}{l_s^2 p^+} \sigma \right)} +
  {\tilde{b}_n}^\dagger e^{-i \left( \omega_n \tau + \frac{n}{l_s^2 p^+}  \sigma \right)}
  \right),
\end{eqnarray}
where

\begin{equation}
R_1^{\mathbf{IIB}} = \frac{R}{\sqrt{2} N_2} \quad \textrm{and} \quad
\omega_n = \sqrt{\mu^2 + \frac{n^2}{l_s^4 (p^+)^2}}.
\end{equation}
The bosonic part of the light-cone Hamiltonian is:

\begin{eqnarray}    \label{eq:IIB2HB}
H^B = \frac{m^2 {R_1^{\mathbf{IIB}}}^2}{2 l_s^4 p^+} + 2\mu N^{(a)}_0 +  2\mu N^{(b)}_0 +
\sum_{n \neq 0}  (\omega_n + \mu) \left( N^{(a)}_n  + N^{(b)}_n \right) +  \nonumber\\ 
+\sum_{n \neq 0}  (\omega_n - \mu) \left(\tilde{N}^{(a)}_n + \tilde{N}^{(b)}_n \right) +  
\sum_{I=5}^8 \sum_{n = -\infty}^{\infty} \omega_n N_n^{I},
\end{eqnarray}
where

\begin{eqnarray}
N^I_n = a_n^I {a_n^I}^\dagger, \quad
N^{(a)}_n = a_n {a_n}^\dagger, \quad
\tilde{N}^{(a)}_n = \tilde{a}_n {\tilde{a}_n}^\dagger, \quad
N^{(b)}_n = b_n {b_n}^\dagger, \quad
\tilde{N}^{(b)}_n = \tilde{b}_n {\tilde{b}_n}^\dagger.
\end{eqnarray}
As expected the Hamiltonian does not depend on $p_1$ and $p_3$, 
with the former being quantized because of the compactness of $x_1$. There is an infinite
degeneracy related to these quantum momenta. In particular, the ground state is characterized by

\begin{equation}        \label{eq:gs2}
\left| k,m,p^3,p^+ \right>,
\end{equation}
where $k$ is defined by $p_1=\frac{k}{R_1^\mathbf{IIB}}$ and $m$ is the winding number.
In order to write down the fermionic part of the Hamiltonian we 
recall that the generalized covariant derivative in the type IIB background is
given by:

\begin{equation}
\mathcal{D}_\beta = \partial_\beta + \frac{1}{4} \partial_{\beta} x^+
 \left[    \omega_{\rho \sigma +} \Gamma^{\rho \sigma} 
         - \frac{1}{2 \cdot 5!} F_{\lambda \nu \rho \sigma \kappa}
          \Gamma^{\lambda \nu \rho \sigma \kappa} i\sigma^2 \Gamma_+   \right]
\end{equation}
and the non-vanishing components of the spin connection and the RR five-form are:

\begin{equation}
\omega_{+12} =\omega_{+34}= -\mu \quad \textrm{and} \quad F_{+1234}=F_{+5678}=2 \mu.
\end{equation}
Using the definition of the eight-component spinor $S^{Ab}$ given previously we get the
fermionic part of the light-cone action:

\begin{equation}
\begin{aligned}
S^F_{lc} = \frac{i}{4 \pi l_s^2 p^+} \int d \tau \int_{0}^{2 \pi l_s^2 p^+} d \sigma
\left[ 
   S^1 \left( \partial_+ - \frac{\mu}{2} \gamma^{12} - \frac{\mu}{2} \gamma^{34} \right) S^1 + 
                                              \right. \\ \left.
   + S^2 \left( \partial_- - \frac{\mu}{2} \gamma^{12} - \frac{\mu}{2} \gamma^{34} \right) S^2 - 
       - 2 \mu  S^1 \gamma^{1234} S^2 
\right].
\end{aligned}
\end{equation}
Finding the mode expansions of $S^{1,2}$ and using it to re-write 
the fermionic part of the Hamiltonian we obtain:

\begin{equation}   \label{eq:IIA2HF}
H^F = \sum_{n=-\infty}^{\infty} S_n^\dagger 
  \left( \omega_n + i \frac{\mu}{2} \gamma^{12} +  i \frac{\mu}{2} \gamma^{12}\right) S_n. 
\end{equation}
Since the eigenvalues of both $i \gamma^{12}$ and $i \gamma^{34}$ are $\pm 1$, each with
multiplicity four, and the two matrices commute, we may choose a suitable basis to arrive at:

\begin{equation}\label{eq:IIB2HF}
H^F = \sum_{n=-\infty}^{\infty}
        \left[
           \sum_{a=1}^{2} \left( \omega_n - \mu \right) F^{(a)}_n +
           \sum_{a=3}^{6}  \omega_n F^{(a)}_n +
           \sum_{a=7}^{8} \left( \omega_n + \mu \right) F^{(a)}_n
        \right].
\end{equation}
Recalling that $\omega_0 = \mu$, we see that 
there are two fermionic zero-modes ($a=1,2$) with vanishing energy.
It means that the ground state (\ref{eq:gs2}) has an additional degeneracy due to the
fermionic degrees of freedom.

\subsection{IIA}

Similarly to the type IIB metric we take the Penrose limit around the geodesic
at $\alpha =\frac{\pi}{4}$. The following coordinate transformation does the job:

\begin{eqnarray}     \label{eq:CT2IIA}
t = \mu x^+ + \frac{2 x^-}{\mu R^2},     \quad
\alpha = \frac{\pi}{4} - \frac{x_4}{R}, \quad
\theta = \mu x^+ + \frac{x_3}{R}, \quad 
\half \phi = \mu x^+  - \frac{x_3}{R},   \quad   \nonumber     \\
\tilde{\theta} =  2 \sqrt{2} \frac{x_2}{R}, \quad                 
Y = \frac{1}{\sqrt{2}} \frac{x_1}{R}, \quad
\rho = \frac{r}{R} \qquad.
\end{eqnarray}
In the $R \to \infty$ limit we find a pp-wave metric with one
"magnetic" term:

\begin{equation}
ds^2 = - 4 dx^+ dx^- -\mu^2 \left(4 x_2^2 + \sum_{I=5}^8 x_I^2 \right) {dx^+}^2 +
 \sum_{i=1}^8 {dx_i}^2 + 4 \mu x_4 dx_3 dx^+,
\end{equation}
where $r^2=\sum_{I=5}^8 x_I^2$ and the other fields are 

\begin{equation}       
H^{(3)}_{+12} = 2 \mu, \quad B^{(2)}_{+1} = 2 \mu x_2,
\quad 
F^{(4)}_{+234} = - 4 \pi \frac{N_1}{R^3} \cdot \frac{2 \sqrt{2}}{4} \mu,
\quad
e^{\Phi} = \sqrt{2} g_s \frac{N_2}{R}.
\end{equation}
and this background also preserves 24 supercharges.
Again, the  $x_1$ direction obeys

\begin{equation}
x_1 \equiv x_1 + 2 \pi \frac{N_2}{\sqrt{2} R}l_s^2
\end{equation}
and this background is a T-dual of the type IIB solution (\ref{eq:IIBm2}).

The calculation of the string spectrum in this background is straightforward. 
The part of the Hamiltonian related to the bosons $x_3$ and $x_4$ is the same as in
the T-dual case, while the part consisting of $x_1$ and $x_2$ appears in  
the type IIA $\alpha=\frac{\pi}{2}$ case. The bosonic part of the Hamiltonian is
equal to:

\begin{eqnarray}    \label{eq:IIA2HB}
H^B = \frac{k^2}{2 {R_1^{\mathbf{IIA}}}^2 p^+} + 2\mu N^{(a)}_0 +  2\mu N^{(b)}_0 +
\sum_{n \neq 0}  (\omega_n + \mu) \left( N^{(a)}_n  + N^{(b)}_n \right) +  \nonumber\\   
+ \sum_{n \neq 0}  (\omega_n - \mu) \left(\tilde{N}^{(a)}_n + \tilde{N}^{(b)}_n \right) +
 \sum_{I=5}^8 \sum_{n = -\infty}^{\infty} \omega_n N_n^{I},
\end{eqnarray}
where

\begin{equation}
R_1^{\mathbf{IIA}} = \frac{N_2}{\sqrt{2} R}l_s^2.
\end{equation}
To find the fermionic spectrum note, that the forms depending part 
of the type IIA covariant derivative is not modified. There is, however, 
a non-vanishing component of the spin connection:

\begin{equation}
\omega_{34+} = - \mu.
\end{equation}
After some algebra one verifies that 
the fermionic part of the Hamiltonian matches the type IIB result (\ref{eq:IIA2HF}).
To summarize, the string spectra in the type IIA and the type IIB backgrounds are the same.

\subsection{The BMN operators}

In our Penrose limit the type IIA dictionary (\ref{eq:dict1})
between the string theory and the QGT is modified according to (\ref{eq:CT2IIA}):

\begin{equation}  \label{eq:dict2IIA}
p^+  =  \frac{\Delta}{\mu R^2}, \quad
H  =  \mu \left( \Delta - J_\theta -2 J_\phi \right)
\quad \textrm{and} \quad
p_3 = -i \partial_{x_3} = \frac{J_\theta - 2 J_\phi}{R}.
\end{equation}
Again, the shift along the compact directions acts on the fields according to (\ref{eq:Shift}).
In the type IIB background the dictionary follows from (\ref{eq:CT2IIB}):

\begin{equation}  \label{dict2IIB}
p^+  =  \frac{\Delta}{\mu R^2}, \quad
H  =  \mu \left( \Delta - i\partial_\chi -  i\partial_\eta -i \partial_\beta \right) \equiv
              \mu \left( \Delta - 2 J \right), \quad
p_3 = \frac{ - i\partial_\chi -  i\partial_\eta + i \partial_\beta}{R} \equiv \frac{J_3}{R}\end{equation}
and

\begin{equation}    \label{eq:p1J1}
p_1 =  -i \partial_{x_1} 
    =\sqrt{2} \frac{  i\partial_\chi -  i\partial_\eta}{R} \equiv \sqrt{2}\frac{2 J_1}{R}
= \frac{k}{R_1^{\mathbf{IIB}}} = \sqrt{2}\frac{k N_2}{R}.
\end{equation}
We recall that the fields $\Phi$,$A$ and $B$ are respectively identified with the coordinates
$z_1$,$z_2$ and $z_3$ defined in (\ref{eq:ABPhiz1z2z3}). One can easily find 
the corresponding eigenvalues of $J_{1,3}$ using these relations.
In particular,  the following identification hold:

\begin{equation}
i \partial_\beta \sim J_\theta \quad \textrm{and}
\quad  i \partial_\chi + i \partial_\eta \sim 2 J_\phi.
\end{equation} 
We summarize the quantum numbers of the bosonic fields in the following table:

\begin{center}
\begin{tabular}{c|c|c|c|c|c|c|c|c|}
          & $\Delta$  & $J_\theta$ &  $J_\phi$  &     $J$   &  $H$   &  $J_1$  & $J_3$ \\
 \hline
         $A_I$  &  1  &     0      &  $\half$   &  $\half$  &  0     & $\half$ &   -1\\  \hline  
         $B_I$  &  1  &     0      &  $\half$   &  $\half$  &  0     &-$\half$ &   -1\\  \hline  
   $\bar{A}_I$  &  1  &     0      &  $-\half$  &  -$\half$ &  2     &-$\half$ &    1\\  \hline 
   $\bar{B}_I$  &  1  &     0      &  $-\half$  &  -$\half$ &  2     & $\half$ &    1\\  \hline  
      $\Phi_I$  &  1  &     1      &     0      &  $\half$  &  0     &  0      &    1\\  \hline
$\bar{\Phi}_I$  &  1  &    -1      &     0      &  -$\half$ &  2     &  0      &    1\\  \hline 
\end{tabular}
\end{center}
As we already mentioned $J_\theta$ and $J_\phi$ are generators of the $SU(2)_R \times U(1)_R$  
supersymmetry group. On other hand, $J_1$ is a generator of an additional $U(1)$ symmetry, 
which is not an $R$-symmetry. It rotates $A$ and $B$
 in opposite directions: $A \to A e^{i \delta}$ and  $B \to B e^{i \delta}$. 
Because of the orbifolding, having $\delta =\frac{2 \pi}{N_2}$ brings us back to the same point.
This agrees with an identification made in (\ref{eq:p1J1}):

\begin{equation}
2 J_1 = k N_2.
\end{equation}
The fermionic sector of the $N=2$ QGT consists of four different four-dimensional spinors:
$\psi_I$ are the gauginos, $\chi_{A_{I}}$ and  $\chi_{B_{I}}$ are the fermionic superpartners of
$A_I$ and $B_I$ in the $\mathcal{N}=1$ hyper-multiplets and  $\psi_{\Phi_{I}}$ 
are the fermionic partners of $\Phi_I$.
The fermionic quantum numbers are:

\begin{center}
\begin{tabular}{c|c|c|c|c|c|c|}
                       &  $\Delta$  & $J_\theta$ &  $J_\phi$   &     $J$      & $H$  \\
 \hline
         $\chi_{A_I}$  &$\frac{3}{2}$&  $\half$   &     0      &    $\half$   & 1    \\  \hline  
         $\chi_{B_I}$  &$\frac{3}{2}$&  $\half$   &     0      &    $\half$   & 1    \\  \hline  
   $\bar{\chi}_{A_I}$  &$\frac{3}{2}$&  $-\half$  &     0      &   $-\half$   & 2    \\  \hline 
   $\bar{\chi}_{B_I}$  &$\frac{3}{2}$&  $-\half$  &     0      &   $-\half$   & 2    \\  \hline  
      $\psi_{\Phi_I}$  &$\frac{3}{2}$&  $\half$   &  $-\half$  &   $-\half$   & 2    \\  \hline
$\bar{\psi}_{\Phi_I}$  &$\frac{3}{2}$&  $-\half$  &  $ \half$  &    $\half$   & 1    \\  \hline 
             $\psi_I$  &$\frac{3}{2}$&  $-\half$  &  $-\half$  &$-\frac{3}{2}$& 3    \\  \hline
       $\bar{\psi}_I$  &$\frac{3}{2}$&  $\half$   &  $ \half$  &$ \frac{3}{2}$& 0    \\  \hline
\end{tabular}
\end{center}
We construct further the BMN operator related to the string theory vacuum state.  
To this end the generalization of (\ref{eq:G}) is required:

\begin{eqnarray}                \label{eq:newG}
&&\mathcal{G}_{K, L, M} \left( A, \Phi, B; \omega^m \right) \equiv  \nonumber\\
&&\frac{1}{\sqrt{(K+L+M)!K!L!M!}} \partial_x^K  \partial_y^L \partial_z^M
  \prod_{s=0}^{K+L+M-1} \left( \omega^{-s/2} x \mathbf{A} + y \mathbf{\Phi}  +
        \omega^{s/2} z \mathbf{B} \right) \vert_{x,y,z=0}.  
\end{eqnarray}
Using this notation a momentum state in the type IIB pp-wave background is related to:
  
\begin{equation}
\left| k, m=0 \right>_{\mathrm{IIB}} \simeq \mathrm{Tr} \left[ 
             \mathcal{G}_{\half (J-J_3) + J_1,J + J_3,\half (J-J_3) - J_1} 
                \left( \mathbf{A}, \mathbf{\Phi}, \mathbf{B}; 1 \right) 
                    \right]
\end{equation}
and we remind that $k = 2\frac{J_1}{N_2}$.
In the type IIA picture this operator describes a winding state:
\begin{equation}
\left| k=0, m \right>_{\mathrm{IIA}} \simeq \mathrm{Tr} \left[ 
            \mathcal{G}_{\half (J-J_3) + \frac{m N_2}{2},J + J_3,\half (J-J_3) - \frac{m N_2}{2}} 
                \left( \mathbf{A}, \mathbf{\Phi}, \mathbf{B}; 1 \right) 
                    \right].
\end{equation}
Similarly, the type IIB winding mode is given by:

\begin{equation}
\left| k=0, m \right>_{\mathrm{IIB}} \simeq \mathrm{Tr} \left[ \mathbf{V}^m
             \mathcal{G}_{\half (J-J_3),J + J_3,\half (J-J_3)} 
                \left( \mathbf{A}, \mathbf{\Phi}, \mathbf{B}; \omega^m \right) 
                    \right].
\end{equation}
One can act on these ground states with two massless fermionic zero-modes 
(see the discussion following (\ref{eq:IIB2HF})).
In view of the fermionic quantum numbers we find that $\psi_I$ is the only fermionic field with
$H^B=0$. Inserting one of the components of $\psi$ into the trace we reproduce an appropriate
 string zero mode state.  

We turn now to the massive zero mode states in the string theory. In the bosonic sector there are
eight massive zero modes: two states created by $a_0^\dagger$ and $b_0^\dagger$ with $H=2 \mu$ and 
four states with $H=\mu$ created by $a_I^\dagger$ ($I=5, \ldots 8$).
The later are given by:

\begin{equation}
a_I^\dagger \left| k, m=0 \right>_{\mathrm{IIB}} \simeq 
  \mathrm{Tr} \left[  Z_i
             \mathcal{G}_{\half (J-J_3) + J_1,J + J_3,\half (J-J_3) - J_1} 
                \left( \mathbf{A}, \mathbf{\Phi}, \mathbf{B}; 1 \right) 
              \right], 
\end{equation}
where 

\begin{equation}
Z_i = D_{i-5} \quad \textrm{for} \quad i= 5, \ldots, 8
\end{equation}
and $D_i$ is understood to act on $A$, $B$ or $\Phi$ to the right of it.
The fifth zero mode is obtained by the insertion of $\mathbf{\Phi}$ into the trace and there is
a perfect agreement between the zero mode energy $H=2 \mu$ and the 
$H = \mu (\Delta - J)$ eigenvalue of the field. 
The last sixth zero mode is realized by acting on the ground state with
an operator ~\cite{Bertolini:2002nr}:

\begin{equation}
\mathcal{D} \sim \bar{\mathbf{A}} \partial_\mathbf{B} - \bar{\mathbf{B}} \partial_\mathbf{A}.
\end{equation}
Namely, one inserts the operator $\bar{\mathbf{A}}\mathbf{B} - \bar{\mathbf{B}}\mathbf{B}$
into the ground state in a totally symmetric way. Again, the energy $H=2 \mu$ matches the 
dimensions of the $\bar{A}$ and $\bar{B}$.

The construction of various non-zero string state by proper insertion of ``impurities''
into the ground state as well as the check of the level matching condition, 
the normalization of the operators and other related issues are quite similar 
to ~\cite{Bertolini:2002nr} and we will not perform it here. Note that provided any operator in
the type IIB description one obtains the type IIA operator by simple replacement of the 
momentum number $k$ and the winding number $m$. 
The string theory prediction of the energy eigenvalues of these operators is given
by (\ref{eq:IIB2HB}) or (\ref{eq:IIA2HB}):

\begin{equation}
\begin{aligned}   
\frac{H^B}{\mu} =& 
       m^2 \frac{ g_{QGT}^2}{2 N_2 J}
         + 2 N^{(a)}_0 +  2 N^{(b)}_0 +
        \sum_{n \neq 0}   
              \left(1 +  \sqrt{1 + n^2\frac{g_{QGT}^2 N_2}{4 J^2}}\right)
                 \left( N^{(a)}_n  + N^{(b)}_n \right) +              \\ 
    &   +\sum_{n \neq 0} 
              \left(-1 +  \sqrt{1 + n^2\frac{g_{QGT}^2 N_2}{4 J^2}}\right) 
                  \left(\tilde{N}^{(a)}_n + \tilde{N}^{(b)}_n \right) +
       \sum_{I=5}^8 \sum_{n = -\infty}^{\infty}  \sqrt{1 + n^2 \frac{g_{QGT}^2 N_2}{4 J^2}} 
                                                                     N_n^{I},          \\ 
\frac{H^F}{\mu} =& 
        \sum_{n=-\infty}^{\infty}
        \Bigg[
           \sum_{a=1}^{2} \left(-1 + \sqrt{1 + n^2 \frac{g_{QGT}^2 N_2}{4 J^2}}\right) F^{(a)}_n +
           \sum_{a=3}^{6} \sqrt{1 + n^2 \frac{g_{QGT}^2 N_2}{4 J^2}} F^{(a)}_n +    \\ 
    &       \sum_{a=7}^{8} \left(1 + \sqrt{1 + n^2 \frac{g_{QGT}^2 N_2}{4 J^2}}\right) F^{(a)}_n
        \Bigg].
\end{aligned}
\end{equation}

\acknowledgments

We would like to thank Jacob Sonnenschein, Yaron Oz, Tadakatsu Sakai 
and Khasanov Oleg for fruitful discussions.
This work was supported in part by the German-Israeli Foundation for
Scientific Research and by the Israel Science Foundation.

\bibliography{p1}

\providecommand{\href}[2]{#2}\begingroup\raggedright\begin{thebibliography}{10}

\bibitem{Berenstein:2002jq}
D.~Berenstein, J.~M. Maldacena, and H.~Nastase, {\it Strings in flat space and
  pp waves from n = 4 super yang mills},  {\em JHEP} {\bf 04} (2002) 013,
  [\href{http://xxx.lanl.gov/abs/hep-th/0202021}{{\tt hep-th/0202021}}].

\bibitem{Blau:2001ne}
M.~Blau, J.~Figueroa-O'Farrill, C.~Hull, and G.~Papadopoulos, {\it A new
  maximally supersymmetric background of iib superstring theory},  {\em JHEP}
  {\bf 01} (2002) 047, [\href{http://xxx.lanl.gov/abs/hep-th/0110242}{{\tt
  hep-th/0110242}}].

\bibitem{Blau:2002dy}
M.~Blau, J.~Figueroa-O'Farrill, C.~Hull, and G.~Papadopoulos, {\it Penrose
  limits and maximal supersymmetry},  {\em Class. Quant. Grav.} {\bf 19} (2002)
  L87--L95, [\href{http://xxx.lanl.gov/abs/hep-th/0201081}{{\tt
  hep-th/0201081}}].

\bibitem{Blau:2002mw}
M.~Blau, J.~Figueroa-O'Farrill, and G.~Papadopoulos, {\it Penrose limits,
  supergravity and brane dynamics},  {\em Class. Quant. Grav.} {\bf 19} (2002)
  4753, [\href{http://xxx.lanl.gov/abs/hep-th/0202111}{{\tt hep-th/0202111}}].

\bibitem{Metsaev:2001bj}
R.~R. Metsaev, {\it Type iib green-schwarz superstring in plane wave ramond-
  ramond background},  {\em Nucl. Phys.} {\bf B625} (2002) 70--96,
  [\href{http://xxx.lanl.gov/abs/hep-th/0112044}{{\tt hep-th/0112044}}].

\bibitem{Itzhaki:2002kh}
N.~Itzhaki, I.~R. Klebanov, and S.~Mukhi, {\it Pp wave limit and enhanced
  supersymmetry in gauge theories},  {\em JHEP} {\bf 03} (2002) 048,
  [\href{http://xxx.lanl.gov/abs/hep-th/0202153}{{\tt hep-th/0202153}}].

\bibitem{Gomis:2002km}
J.~Gomis and H.~Ooguri, {\it Penrose limit of n = 1 gauge theories},  {\em
  Nucl. Phys.} {\bf B635} (2002) 106--126,
  [\href{http://xxx.lanl.gov/abs/hep-th/0202157}{{\tt hep-th/0202157}}].

\bibitem{PandoZayas:2002rx}
L.~A. Pando~Zayas and J.~Sonnenschein, {\it On penrose limits and gauge
  theories},  {\em JHEP} {\bf 05} (2002) 010,
  [\href{http://xxx.lanl.gov/abs/hep-th/0202186}{{\tt hep-th/0202186}}].

\bibitem{Corrado:2002wi}
R.~Corrado, N.~Halmagyi, K.~D. Kennaway, and N.~P. Warner, {\it Penrose limits
  of rg fixed points and pp-waves with background fluxes},
  \href{http://xxx.lanl.gov/abs/hep-th/0205314}{{\tt hep-th/0205314}}.

\bibitem{Hubeny:2002vf}
V.~E. Hubeny, M.~Rangamani, and E.~Verlinde, {\it Penrose limits and non-local
  theories},  {\em JHEP} {\bf 10} (2002) 020,
  [\href{http://xxx.lanl.gov/abs/hep-th/0205258}{{\tt hep-th/0205258}}].

\bibitem{Gimon:2002sf}
E.~G. Gimon, L.~A. Pando~Zayas, and J.~Sonnenschein, {\it Penrose limits and rg
  flows},  \href{http://xxx.lanl.gov/abs/hep-th/0206033}{{\tt hep-th/0206033}}.

\bibitem{Brecher:2002ar}
D.~Brecher, C.~V. Johnson, K.~J. Lovis, and R.~C. Myers, {\it Penrose limits,
  deformed pp-waves and the string duals of n = 1 large n gauge theory},  {\em
  JHEP} {\bf 10} (2002) 008,
  [\href{http://xxx.lanl.gov/abs/hep-th/0206045}{{\tt hep-th/0206045}}].

\bibitem{Oz:2002ku}
Y.~Oz and T.~Sakai, {\it Penrose limit and six-dimensional gauge theories},
  {\em Phys. Lett.} {\bf B544} (2002) 321--328,
  [\href{http://xxx.lanl.gov/abs/hep-th/0207223}{{\tt hep-th/0207223}}].

\bibitem{Bhattacharya:2002zf}
S.~Bhattacharya and S.~Roy, {\it Penrose limit and ncym theories in diverse
  dimensions},  \href{http://xxx.lanl.gov/abs/hep-th/0209054}{{\tt
  hep-th/0209054}}.

\bibitem{Bhattacharya:2002qx}
S.~Bhattacharya and S.~Roy, {\it More on penrose limits and non-local
  theories},  {\em JHEP} {\bf 01} (2003) 064,
  [\href{http://xxx.lanl.gov/abs/hep-th/0210072}{{\tt hep-th/0210072}}].

\bibitem{Alishahiha:2002ev}
M.~Alishahiha and M.~M. Sheikh-Jabbari, {\it The pp-wave limits of orbifolded
  ads(5) x s(5)},  {\em Phys. Lett.} {\bf B535} (2002) 328--336,
  [\href{http://xxx.lanl.gov/abs/hep-th/0203018}{{\tt hep-th/0203018}}].

\bibitem{Mukhi:2002ck}
S.~Mukhi, M.~Rangamani, and E.~Verlinde, {\it Strings from quivers, membranes
  from moose},  {\em JHEP} {\bf 05} (2002) 023,
  [\href{http://xxx.lanl.gov/abs/hep-th/0204147}{{\tt hep-th/0204147}}].

\bibitem{Alishahiha:2002jj}
M.~Alishahiha and M.~M. Sheikh-Jabbari, {\it Strings in pp-waves and worldsheet
  deconstruction},  {\em Phys. Lett.} {\bf B538} (2002) 180--188,
  [\href{http://xxx.lanl.gov/abs/hep-th/0204174}{{\tt hep-th/0204174}}].

\bibitem{Naculich:2002fh}
S.~G. Naculich, H.~J. Schnitzer, and N.~Wyllard, {\it pp-wave limits and
  orientifolds},  {\em Nucl. Phys.} {\bf B650} (2003) 43--74,
  [\href{http://xxx.lanl.gov/abs/hep-th/0206094}{{\tt hep-th/0206094}}].

\bibitem{Bertolini:2002nr}
M.~Bertolini, J.~de~Boer, T.~Harmark, E.~Imeroni, and N.~A. Obers, {\it Gauge
  theory description of compactified pp-waves},  {\em JHEP} {\bf 01} (2003)
  016, [\href{http://xxx.lanl.gov/abs/hep-th/0209201}{{\tt hep-th/0209201}}].

\bibitem{Douglas:1996sw}
M.~R. Douglas and G.~W. Moore, {\it D-branes, quivers, and ale instantons},
  \href{http://xxx.lanl.gov/abs/hep-th/9603167}{{\tt hep-th/9603167}}.

\bibitem{Michelson:2002wa}
J.~Michelson, {\it (twisted) toroidal compactification of pp-waves},  {\em
  Phys. Rev.} {\bf D66} (2002) 066002,
  [\href{http://xxx.lanl.gov/abs/hep-th/0203140}{{\tt hep-th/0203140}}].

\bibitem{Alishahiha:2002nf}
M.~Alishahiha, M.~A. Ganjali, A.~Ghodsi, and S.~Parvizi, {\it On type iia
  string theory on the pp-wave background},
  \href{http://xxx.lanl.gov/abs/hep-th/0207037}{{\tt hep-th/0207037}}.

\bibitem{Mizoguchi:2003be}
S.~Mizoguchi, T.~Mogami, and Y.~Satoh, {\it A note on t-duality of strings in
  plane-wave backgrounds},  \href{http://xxx.lanl.gov/abs/hep-th/0302020}{{\tt
  hep-th/0302020}}.

\bibitem{Fayyazuddin:1999zu}
A.~Fayyazuddin and D.~J. Smith, {\it Localized intersections of m5-branes and
  four-dimensional superconformal field theories},  {\em JHEP} {\bf 04} (1999)
  030, [\href{http://xxx.lanl.gov/abs/hep-th/9902210}{{\tt hep-th/9902210}}].

\bibitem{Gueven:2000ru}
R.~Gueven, {\it Plane wave limits and t-duality},  {\em Phys. Lett.} {\bf B482}
  (2000) 255--263, [\href{http://xxx.lanl.gov/abs/hep-th/0005061}{{\tt
  hep-th/0005061}}].

\bibitem{Bena:2002kq}
I.~Bena and R.~Roiban, {\it Supergravity pp-wave solutions with 28 and 24
  supercharges},  \href{http://xxx.lanl.gov/abs/hep-th/0206195}{{\tt
  hep-th/0206195}}.

\bibitem{Metsaev:2002re}
R.~R. Metsaev and A.~A. Tseytlin, {\it Exactly solvable model of superstring in
  plane wave ramond- ramond background},  {\em Phys. Rev.} {\bf D65} (2002)
  126004, [\href{http://xxx.lanl.gov/abs/hep-th/0202109}{{\tt
  hep-th/0202109}}].

\end{thebibliography}\endgroup

\end{document}